# Cryo-EM structures of atomic surfaces and host-guest chemistry in metal-organic frameworks

**Authors:** Yuzhang Li[1,†], Kecheng Wang[1,†], Weijiang Zhou[2,†], Yanbin Li[1], Rafael Vila[1], William Huang[1], Hongxia Wang[1], Guangxu Chen[1], Gong-Her Wu[3], Yuchi Tsao[4], Hansen Wang[1], Robert Sinclair[1], Wah Chiu[2,3,5,6], and Yi Cui[1,7*]

**Affiliations:**

[1]Department of Materials Science and Engineering, Stanford University, Stanford, California 94305, USA.

[2]Biophysics Program, School of Medicine, Stanford University, Stanford, California 94305, USA.

[3]Department of Bioengineering, Stanford University, Stanford, California 94305, USA.

[4]Department of Chemistry, Stanford University, Stanford, California 94305, USA.

[5]Department of Microbiology and Immunology, Stanford University School of Medicine, Stanford, California 94305, USA.

[6]Division of CryoEM and Bioimaging, SSRL, SLAC National Accelerator Laboratory, Menlo Park, California 94025, USA.

[7]Stanford Institute for Materials and Energy Sciences, SLAC National Accelerator Laboratory, Menlo Park, California 94025, USA.

[*]Correspondence to: yicui@stanford.edu

[†]These authors contributed equally to this work.

**Abstract**:

Host-guest interactions govern the chemistry of a broad range of functional materials. However, the weak bonding between host and guest prevents atomic-resolution studies of their structure and chemistry using transmission electron microscopy (TEM). This problem is exacerbated in metal-organic frameworks (MOF), in which the host framework is easily damaged by the electron beam. Here, we use cryogenic-electron microscopy (cryo-EM) to simultaneously address these two challenges and resolve the atomic surface structure of zeolitic imidazolate framework (ZIF-8) and its interaction with guest $CO_2$ molecules. We image atomic step-edge sites on the ZIF-8 surface that provides possible insight to its growth behavior. Furthermore, we observe two distinct binding sites for $CO_2$ within the ZIF-8 pore, which are both predicted by density functional theory (DFT) to be energetically favorable. This $CO_2$ insertion induces an apparent ~3% lattice expansion along the <002> and <011> directions of the ZIF-8 unit cell. The ability to stabilize MOFs and preserve their host-guest chemistry opens a rich materials space for scientific exploration and discovery using cryo-EM.

**Main Text**:

Metal-organic frameworks (MOFs) are a large class of highly porous materials[1–3] whose chemistry and crystalline structure can be tuned for potential applications in gas storage[4–6], separations[7], and catalysis[8]. Interactions between the host framework and guest molecule are central to such applications but are poorly understood at the single particle level. Although transmission electron microscopy (TEM) has been used to study the empty MOF framework[9–13], guest insertion within the MOFs could not be imaged. The weak bonding between the guest molecule and framework is easily damaged even under low electron doses[14,15]. Furthermore, guest molecules are likely to desorb from MOF pores under the high vacuum condition (~$10^{-6}$ mbar) of the TEM at room temperature (supplementary text). Therefore, current studies rely on ensemble measurements using X-ray/neutron diffraction[16,17], nuclear magnetic resonance[18], or theoretical simulations[19]. However, the structural information obtained by these methods is averaged over bulk particles. Direct observations of individual MOF particles and their interaction with guest species at high spatial resolution have not been possible.

Recently, cryogenic-electron microscopy (cryo-EM) was shown to be a powerful tool beyond structural biology. For instance, reactive lithium battery materials have been successfully stabilized for imaging and spectroscopy[20–24]. Such cryogenic condition may also reduce radiation damage to the MOF framework (supplementary text). However, cryo-EM procedures developed for biological or battery materials are not necessarily compatible with MOFs. Here, we establish a new cryo-EM protocol to reveal atomic host-guest structures within MOFs, demonstrating that these entities, held together by weak interactions, can be preserved for high-resolution imaging at cryogenic condition (Fig. 1). As an example of this general approach, we investigate the surface structure of a zeolitic imidazolate framework (ZIF-8) and its host-guest chemistry with $CO_2$ molecules, with potential implications for both carbon capture and gas separations[25].

ZIF-8 is a body-centered cubic crystal that has a sodalite topology (space group $I\bar{4}3m$). The ZIF-8 particles (~100 nm in diameter) synthesized in this study (fig. S2) are confirmed to be highly crystalline via X-ray diffraction (XRD; fig. S1). Diffraction contrast from the ZIF-8 particles observed in low-magnification TEM images (Fig. 2a) further suggests a crystalline specimen. However, the bonding between the inorganic metal centers ($Zn^{2+}$) and the organic linkers (2-methylimidazole) of the ZIF-8 framework is extremely sensitive to high electron doses. After exposure to ~50 $e^-/Å^2$ at room temperature, the ZIF-8 particle quickly becomes amorphized, as indicated by both the high-resolution TEM (HRTEM) image (Fig. 2b) and its fast Fourier transform (FFT; Fig. 2b inset). Indeed, previous studies[12,13] have shown that MOF materials quickly lose their crystallinity with an accumulated electron dose of only ~10-20 $e^-/Å^2$, with ZIF-8 becoming completely amorphous after 70 $e^-/Å^2$ at room temperature. Consequently, any existing host-guest structure would be difficult to identify from the amorphized particle. To overcome these challenges, we established a flash-freezing protocol modified from cryo-EM methodologies used in structural biology[26] to preserve the MOFs at their operating environment (e.g. empty or filled) for high-resolution imaging. First, the as-synthesized ZIF-8 particles were dispersed onto a carbon support TEM grid and vacuum-dried to remove all solvent from the ZIF-8 pore space. Samples were then directly plunged into liquid nitrogen while maintaining vacuum or kept in a 1 bar pressure $CO_2$ environment (Fig. 1a, methods). This freezing process prevents any air exposure and kinetically inhibits $CO_2$ desorption from the ZIF-8, preserving the host-guest interactions that were present at room temperature (fig. S3).

Cryogenic temperature is known to reduce radiation damage induced by the electron beam[14,27]. In our cryo-EM imaging experiment, we keep the specimen at −170 °C and expose it to different cumulative electron doses to determine its radiation damage sensitivity (fig. S8). We use a magnification corresponding to a pixel size of 0.68 Å by 0.68 Å to facilitate imaging at atomic resolution. In addition, a direct-detection electron-counting camera has a high quantum efficiency to enable acquisition of images with high signal-to-noise at all frequencies and a high frame rate to allow recording of multiple frames per specimen area followed by subsequent frame alignment to minimize beam-induced drift[28].

Figure 2c is a typical cryo-EM image of a ZIF-8 particle that was held under vacuum before flash-freezing. When viewed along the <111> zone axis, the ZIF-8 rhombic dodecahedral particle is hexagonal (dashed white lines) with well-defined {011} edge planes joined by sharp vertices (fig. S4). Furthermore, the crystalline structure of the ZIF-8 particle is clearly preserved. Under low dose and low temperature conditions, the ZIF-8 structure shows structure information down to 1.86 Å as evidenced by the FFT of this image (Fig. 2d, bottom), exceeding previously reported resolutions of ZIF-8. A simulated TEM image (Fig. 2d inset) also shows good agreement with the raw image. Even after exposure to 25 $e^-/Å^2$ that would ordinarily result in significant damage to the framework, ZIF-8 remains pristine at low temperatures (Fig. 2d). When the number of frames taken and the cumulative exposure are increased (Fig. 2e, f), ZIF-8 shows only partial loss of crystallinity after 90 $e^-/Å^2$ at low temperature, while the same specimen becomes fully amorphous after exposure of 50 $e^-/Å^2$ at room temperature (Fig. 2b). Quantitatively, the plot of normalized intensity in the FFT as a function of cumulative electron exposure (fig. S8) demonstrates that high-resolution information (3 Å) is still retained after 90 e$^-$/Å$^2$. These experiments and the corresponding raw, unprocessed images (fig. S6) establish the increased stability of ZIF-8 using cryo-EM, allowing further exploration into its surface atomic structure and host-guest chemistry.

The atomic surfaces of materials often provide insight into their growth mechanism. In particular, the shape and size uniformity of ZIF-8 particles has been attributed to the formation of surface steps[29,30]. Unfortunately, surface structures and their possible defects are more sensitive than the bulk to the electron beam, making atomic-resolution imaging difficult even at low electron doses[12]. To preserve and study such surface structures immediately after growth, the as-synthesized ZIF-8 particles were flash-frozen and imaged without vacuum-drying (methods). Using ~7 e$^-$/Å$^2$ cumulative electron dose, we have made a number of structure feature observations which were previously not seen at room temperature. Figure 3a is a denoised[31] cryo-EM image of ZIF-8 after synthesis, which appears to be free of defects in the bulk and directly matches the ZIF-8 crystal structure projected along the <111> direction. At a positive defocus of 250 nm, the hexagonal columns of bright contrast correspond to the "Zn clusters" centers of the framework, as confirmed by TEM simulation (methods). Here, we denote a "Zn cluster" as the collection of Zn metal centers and connecting ligands that comprise one of the six ZIF-8 hexagonal vertices visible along the <111> projection (Figure 3b inset). With cryo-EM, it is now possible to resolve atomically sharp surfaces of ZIF-8, demonstrating the striking resolution and stability afforded by this technique. We find that the exposed {011} ZIF-8 surfaces terminate with doubly-coordinated Zn clusters, indicating that Zn clusters on the surface are joined with two other clusters. By reducing the number of dangling ligands, this doubly-coordinated Zn cluster conformation exposes a thermodynamically stable termination surface. Interestingly, these surfaces are not atomically pristine. Instead, step-edge defects are clearly present in regions I (Fig. 3b) and II (Fig. 3c), which may play an important role in the growth mechanism of ZIF-8. Particle growth of ZIF-8 requires the addition of Zn clusters to the exposed surface. This process is not random. Whereas an adatom cluster of Zn at a surface site would only be single-coordinated, addition at a step-edge site would enable double-coordination of the new Zn cluster (Fig. 3d). Therefore, it is thermodynamically favorable for ZIF-8 growth to initiate at step-edge sites rather than surface sites (Fig. 3e), which is consistent with our observations (fig. S5, S7). Finally, bright intensity is visible inside the 6-ring channels of ZIF-8 (Fig. 3b, c), which may come from the solvent molecule within the pore that exists during synthesis (i.e. acetone or methanol). Along with this observation, the low vapor pressure of guest molecules at cryogenic temperature ($\sim 10^{-10}$ mbar for $CO_2$) suggests that host-guest interactions can be stabilized for imaging using cryo-EM (supplementary text).

Raw, unprocessed cryo-EM images of an empty (Fig. 4a) and $CO_2$-filled (Fig. 4d) ZIF-8 particle appear to be very similar when taken at a defocus value of –225 nm. To make a fair comparison between these two images, we correct the "contrast inversion" (methods) introduced by the contrast transfer function (CTF) of the objective lens[32]. Here, the weak-phase object approximation applies to sample thicknesses up to ~100 nm owing to the low density of ZIF-8, which reduces its effective scattering thickness[13]. This standard image processing procedure (methods) generates CTF-corrected images of empty ZIF-8 (Fig. 4b) and $CO_2$-filled ZIF-8 (Fig. 4e) in which the regions of bright intensity represent mass density[33,34], making interpretation of such CTF-corrected images more straightforward. Zn clusters at the unit cell vertices appear as bright dots, forming the expected hexagonal honeycomb lattice along the <111> projection in both the empty and filled ZIF-8 particle. Within the ZIF-8 pore, distinguishing features between the empty and filled state are quite dramatic. Whereas density is absent within the center pore space of empty ZIF-8 unit cells, bright contrast is clearly visible in the middle of the $CO_2$-filled ZIF-8 pore. This suggests that $CO_2$ adsorption is centered within the 6-ring window along the <111> direction. Indeed, TEM simulations (fig. S9, S10) and density-functional theory (DFT) calculations of favorable $CO_2$ binding sites in ZIF-8 support our observations[35]. Using such calculations, we simulate the structure of a ZIF-8 unit cell with the DFT-optimized binding location of $CO_2$ molecules (simplified as red spheres) projected along the <111>. The simulated structure (Fig. 4g) matches well with the CTF-corrected cryo-EM image (Fig. 4e), providing strong evidence that $CO_2$ molecules are successfully preserved within the framework at cryogenic conditions and can be directly imaged using cryo-EM.

Close inspection of a single ZIF-8 unit cell reveals several key features of the host material during guest insertion. In the empty ZIF-8 pore (Fig. 4c), there are three areas of density (circled in blue) just inside the edges of the unit cell. This contrast likely comes from the three organic imidazolate linkers that protrude into the pore space, forming the 6-ring windows that connect each individual pore. Interestingly, these moieties are not observed in ZIF-8 unit cells after $CO_2$ loading (Fig. 4f), suggesting that the imidazolate linkers can rotate out of view to open the narrow 6-ring window (3.4 Å) for $CO_2$ (3.3 Å kinetic diameter) adsorption and diffusion throughout the framework. These results support previous descriptions of a "gate opening" phenomenon in ZIF-8, which explain how molecules larger than the 6-ring window (e.g. $N_2$ 3.6 Å kinetic diameter) can be accommodated into ZIF-8 pores via linker rotation[36–39]. Although it is generally accepted that flexibility of the 6-ring windows occurs at various loading conditions, the ZIF-8 pore cavity itself is largely thought to remain rigid upon $CO_2$ insertion at ambient pressures. Surprisingly, we discover a 3% expansion of the <011> lattice repeat during $CO_2$ loading at atmospheric pressure. Figure 4h plots the integrated pixel intensities for the empty and $CO_2$-filled ZIF-8 unit cells along the <011> direction. The <011> lattice spacing for the empty ZIF-8 averaged over 10 unit cells is measured to be 12.04 Å (fig. S11), which matches well with the experimental value of 12.03 Å from XRD measurement[36]. The precision in this TEM lattice measurement is approximately 0.07 Å (methods). After $CO_2$ loading, the ZIF-8 <011> lattice expands to 12.34 Å (fig. S12). Though lattice expansions of this magnitude (~3%) are possible for ZIF-8 (ref. 40), they have predominantly been observed at high pressure loadings (~$10^4$ bar) of non-interacting gas molecules (e.g. $N_2$, $O_2$, Ar). However, the $CO_2$ still induces a similar unit cell expansion at much lower pressure (~1 bar $CO_2$ loading) in our experiments. This observation implies a strong interaction between $CO_2$ and the ZIF-8 framework, which further suggests ZIF-8 might be a promising carbon capture material.

To reinforce our results, it is important to investigate the host-guest structures along more than one zone axis. Projected in the <001> direction, the rhombic dodecahedral ZIF-8 particle appears as a square (Fig. 5a, 5d). With CTF-correction to address the contrast inversion issue mentioned above, the cryo-EM images of empty (Fig. 5b) and $CO_2$-filled (Fig. 5e) ZIF-8 can be more easily interpreted. Contrast is strongest for the Zn clusters, which are located at the bright square spots. These clusters are connected by the imidazolate linkers that are visible as lines of white contrast. When viewed along the <001> direction (Fig. 5c), ZIF-8 displays two structurally distinct pore cavities: one is located at the vertex of the Zn cluster (circled in green) and the second is adjacent to the square edge of the Zn cluster (boxed in orange). Both cavities do not appear to contain any density in the empty ZIF-8 (Fig. 5c). Upon $CO_2$ adsorption, bright contrast can be observed in the center of the 4-ring window (Fig. 5f). Interestingly, this appears only in the pore cavity located at the vertex of the Zn cluster and not the cavity at the Zn cluster edge. Although adsorption at the center of these particular 4-ring windows is different from the $CO_2$ binding location observed along the <111> direction, DFT calculations show that $CO_2$ binding centered around the 4-ring window is also energetically favorable[35]. The simulated ZIF-8 structure with the predicted $CO_2$ adsorption site (Fig. 5g) closely resembles the cryo-EM image (Fig. 5f), indicating that more than one adsorption site exists. Furthermore, the lattice expansion observed for the <011> direction is also present along the <002> direction (Fig. 5h). Whereas an 8.56 Å <002> lattice spacing is measured for the empty ZIF-8 (fig. S13), the <002> expands 3% to 8.78 Å after $CO_2$ insertion (fig. S14). The similar unit cell expansion and additional adsorption sites revealed here demonstrate the increased information gained by simply observing the particles along different crystallographic directions. The apparent rotation of the imidazolate linkers and expansion of the ZIF-8 cavity observed in this study have important implications for gas diffusivity throughout the porous network, which would fundamentally impact ZIF-8 performance as a carbon capture and gas separation material.

Our work here demonstrates the powerful utility for cryo-EM to preserve and image atomic MOF surface structures and guest molecules within its pore cavities. We revealed that ZIF-8 growth likely initiates at a step-edge surface site and discovered structural changes in ZIF-8 during $CO_2$ insertion. This study opens opportunities to further investigate such host-guest interactions to develop a complete picture of MOF adsorption kinetics at the single particle level. Using cryo-EM, the many applications of different MOFs with distinct guest molecules can be probed at the atomic level. New discoveries and findings will provide insight to their fundamental operation and shape future designs of these tunable materials.

**Acknowledgements**

The authors thank Dave Bushnell and Dong-Hua Chen for training at the Stanford-SLAC cryo-EM facilities. Yuzhang Li acknowledges the Intelligence Community Fellowship for funding. Part of this work was performed at the Stanford Nano Shared Facilities (SNSF), supported by the National Science Foundation under award ECCS-1542152. R.S. acknowledges support from the National Institutes of Health (NIH) under grant no. U54 CA199075. W.C. acknowledges support by the NIH under grant nos. P41GM103832 and S10OD021600. Y.C. acknowledges support from the Assistant Secretary for Energy Efficiency and Renewable Energy, Office of Vehicle Technologies, Battery Materials Research (BMR) and Battery 500 Program of the U.S. Department of Energy.

**Author contributions**

Yuzhang Li, and Y.C. conceived the idea and designed the experiments. Yuzhang Li, Yanbin Li, H.W., and G.C. established the cryo-plunging methodologies. Yuzhang Li, K.W., and Y.T. synthesized the ZIF-8 particles. Yuzhang Li, Yanbin Li, G.H.W. and W.Z. performed the cryo-EM experiments. W.Z. processed the CTF-corrected images. R.V. simulated the TEM images according to experimental imaging conditions. W.H., and H.W. conducted other electron microscopy characterizations at room temperature. Yuzhang Li, K.W., W.Z., R.S., W.C. and Y.C. interpreted the results and co-wrote the paper. All authors discussed the results and commented on the manuscript.

**Competing financial interests**

The authors declare no competing financial interests.

**Figures Captions**

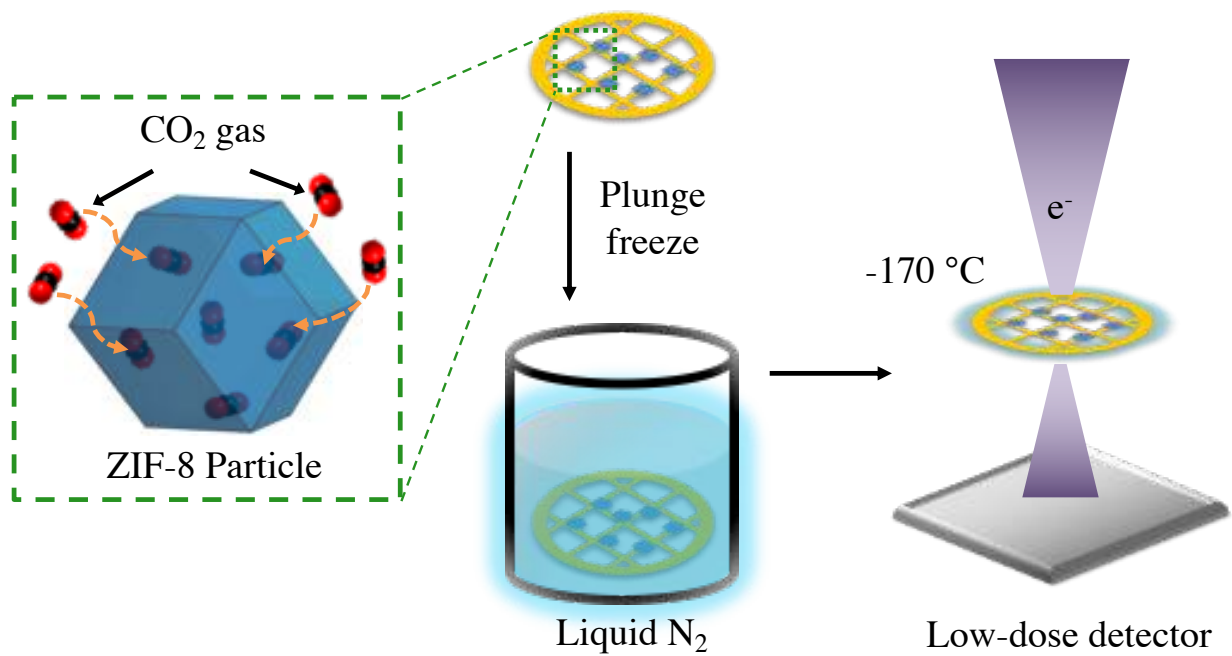

**Figure 1 | Preserving and stabilizing host-guest interactions using Cryo-EM.** Vacuum-dried ZIF-8 particles are exposed to $CO_2$ gas at ambient temperature and pressure after synthesis. In this environment, the particles are then plunge-frozen directly into liquid nitrogen to freeze in the host-guest structure and chemistry. Low-dose images are then recorded at cryogenic condition using a direct electron detector.

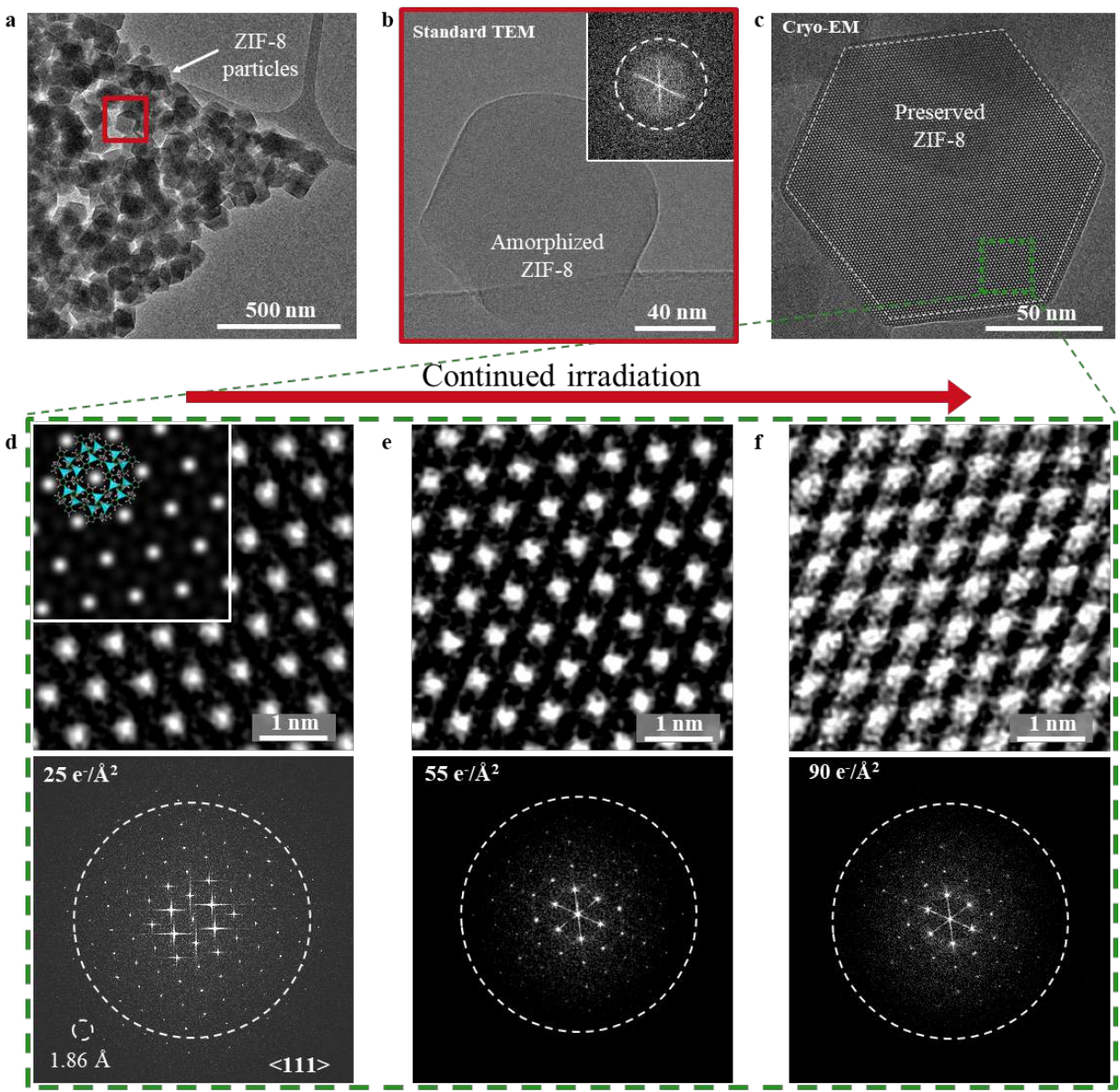

**Figure 2 | Electron irradiation of ZIF-8 at cryogenic temperatures. a**, TEM image of ZIF-8 particles taken at room temperature with electron dose rate 2 e$^-$/Å$^2$/s for ~1s. Diffraction contrast suggests sample is crystalline. **b**, HRTEM image of red boxed region from (**a**). After accumulated electron dose of ~50 e$^-$/Å$^2$, ZIF-8 becomes completely amorphous. Inset: corresponding FFT showing no crystallinity in the sample. **c**, Cryo-EM image of ZIF-8 (outlined by dashed white line) taken along the <111> direction at –170 °C with electron dose rate of ~4.5 e$^-$/Å$^2$/s for 1.5s. At –225 nm defocus, bright spots in the ZIF-8 lattice correspond to empty pore space. **d**, **e**, **f**, Magnified images of green boxed region from (**c**) exposed to electron doses of 25 e$^-$/Å$^2$ (**d**), 50 e$^-$/Å$^2$ (**e**), and 90 e$^-$/Å$^2$ (**f**) with corresponding FFT pattern below. The dashed circle on the FFT represents an information transfer of 2.5 Å. Information transfer of 1.86 Å is possible, as indicated by the reflection circled in (**d**). Inset of (**d**): The ZIF-8 atomic structure is overlaid onto the simulated TEM image, which is calculated using –225 nm defocus and 100 nm sample thickness. The simulated image matches reasonably well with the experimental cryo-EM image.

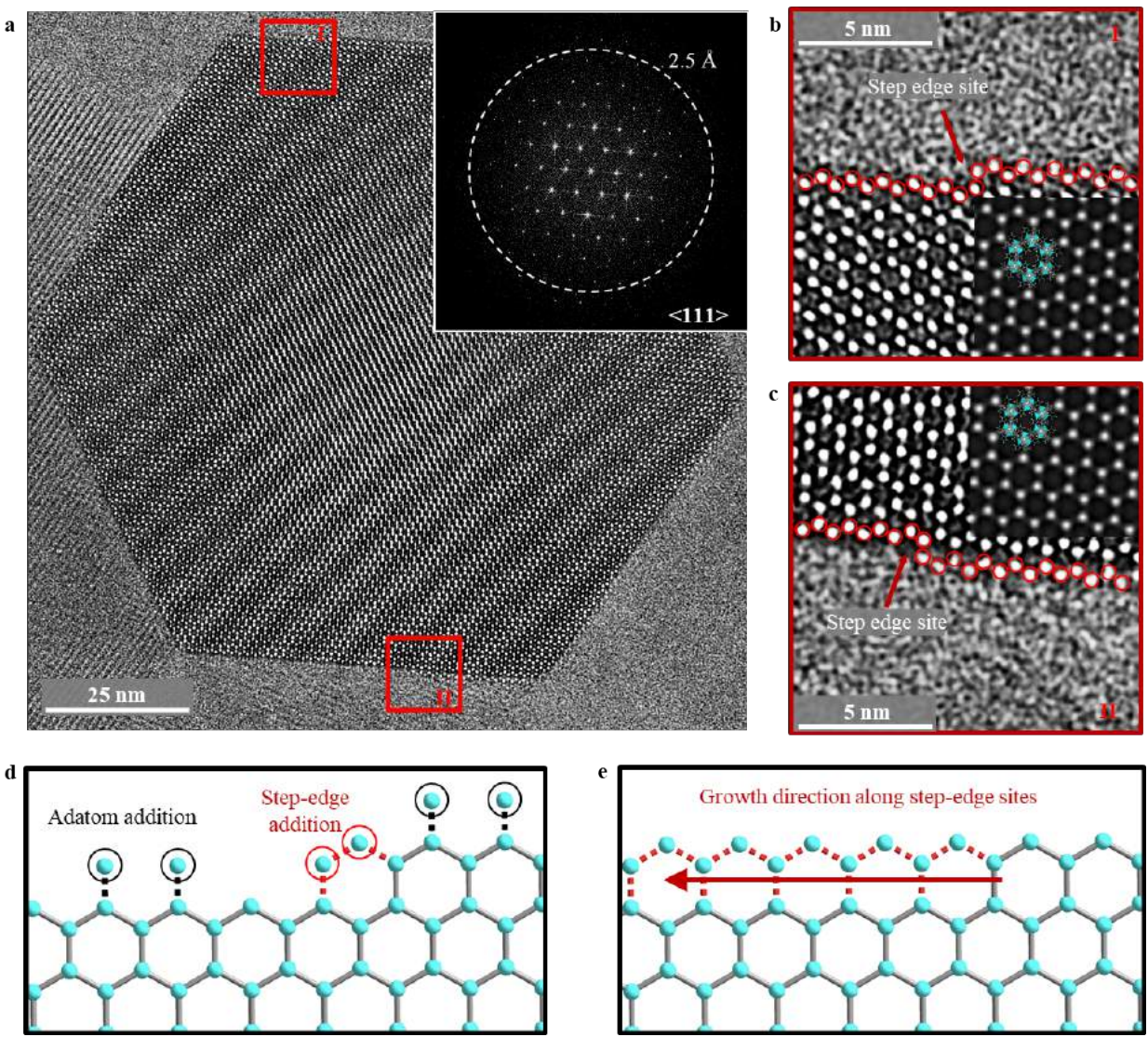

**Figure 3 | Atomic surface and step-edge sites of ZIF-8 particles. a**, Denoised cryo-EM image of ZIF-8 after synthesis (not vacuum-dried) taken at positive 250 nm defocus with electron dose rate of ~4.5 e-/Å$^2$/s for 1.5s. Bright spots correspond to Zn clusters. Inset: corresponding FFT pattern. White dashed circle represents information transfer of 2.5 Å. **b**, **c**, Magnified images of region I (**b**) and region II (**c**) boxed in red from (**a**). The ZIF-8 atomic structure is overlaid onto a simulated TEM image in the second inset, which is calculated using positive 250 nm defocus and 100 nm sample thickness. The simulated TEM image matches well with the experimental cryo-EM image. The Zn clusters on the surface are double-coordinated and are outlined in red circles. A step-edge site is observed in both (**b**) and (**c**), indicated by the red arrow. **d**, Schematic of possible surface additions of Zn clusters during ZIF-8 growth. Whereas adatom addition (circled in black) would result in single-coordinated Zn clusters, addition at the step-edge site (circled in red) would result in double-coordinated Zn clusters. **e**, Schematic of ZIF-8 growth initiated at the step-edge site, which is thermodynamically more favorable.

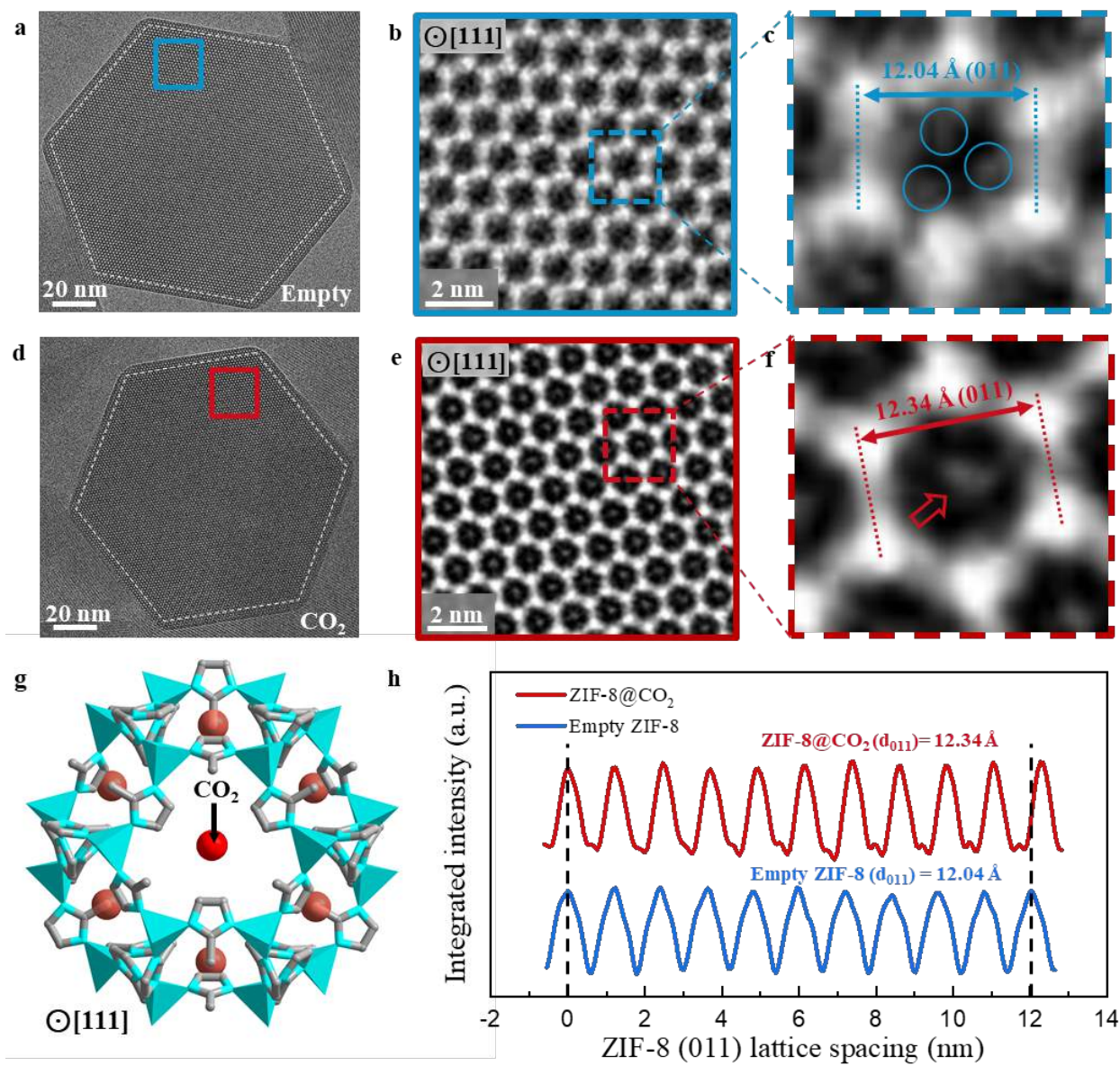

**Figure 4 | Host-guest structures within ZIF-8 viewed along <111> projection. a**, Cryo-EM image of vacuum-dried, empty ZIF-8 (outlined by white dashed lines) taken with electron dose rate of ~4.5 e⁻/Å²/s for 1.5s along the <111> projection. **b**, CTF-corrected denoised image of blue boxed region from (**a**). Bright regions correspond to mass density. **c**, Magnified image of a single ZIF-8 unit cell from (**b**). Density (circled in blue) near the interior edge of the unit cell may correspond to the organic imidazolate linkers. **d**, Cryo-EM image of $CO_2$-filled ZIF-8 particle (outlined by white dashed lines) taken with electron dose rate of ~4.5 e⁻/Å²/s for 1.5s along the <111> projection. **e**, CTF-corrected denoised image of red boxed region from (**d**). Bright regions correspond to mass density. Contrast in the center of the 6-ring window is clearly observed for multiple unit cells. **f**, Magnified image of a single ZIF-8 unit cell from (**e**). Density at the center of the unit cell (indicated by red arrow) likely corresponds to $CO_2$ adsorbed within ZIF-8. **g**, Simulated structure of ZIF-8 with DFT-predicted binding site of $CO_2$ (simplified as red spheres) along the <111> projection. This is consistent with the experimental cryo-EM image from (**f**). **h**,

Integrated intensity of ZIF-8 plotted over 10 unit cells along the <011> direction, indicating a 3% lattice expansion when $CO_2$ is introduced to ZIF-8.

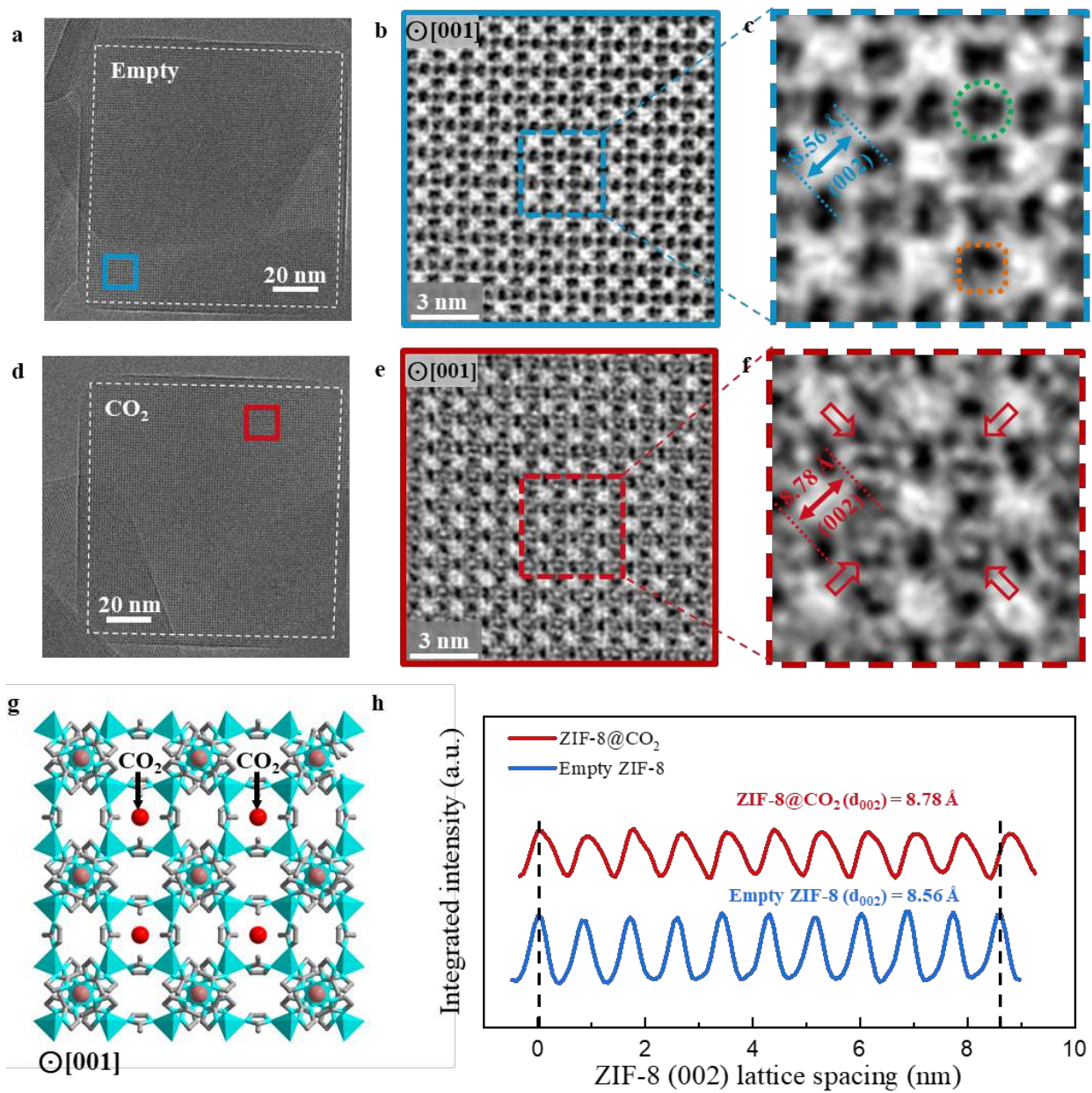

**Figure 5 | Host-guest structures within ZIF-8 viewed along <001> projection. a**, Cryo-EM image of vacuum-dried, empty ZIF-8 (outlined by white dashed lines) taken with electron dose rate of ~4.5 e$^-$/Å$^2$/s for 1.5s along the <001> projection. **b**, CTF-corrected denoised image of blue boxed region from (**a**). Bright regions correspond to mass density. **c**, Magnified image of ZIF-8 unit cells in blue box from (**b**). Green circle indicates pore cavity at the vertices of the Zn clusters. Orange square indicates pore cavity at the edges of the Zn clusters. **d**, Cryo-EM image of $CO_2$-filled ZIF-8 particle (outlined by white dashed lines) taken with electron dose rate of ~4.5 e$^-$/Å$^2$/s for 1.5s along the <001> projection. **e**, CTF-corrected denoised image of red boxed region from (**d**). Bright regions correspond to mass density. Contrast near the center of the 4-ring window is clearly observed for multiple unit cells. **f**, Magnified image of ZIF-8 unit cells in red box from (**e**). Density near the center of the unit cell (indicated by red arrows) likely corresponds to $CO_2$

adsorbed within ZIF-8. Note that only the pore cavity at the vertices of the 4-ring window contain the density in the center. **g**, Simulated structure of ZIF-8 with DFT-predicted binding site of $CO_2$ (simplified as red spheres) along the <001> projection. This is consistent with the experimental cryo-EM image from (**f**). **h**, Integrated intensity of ZIF-8 plotted over 10 unit cells along the <002> direction, indicating a 3% lattice expansion when $CO_2$ is introduced to ZIF-8.

**Supplementary Figures**

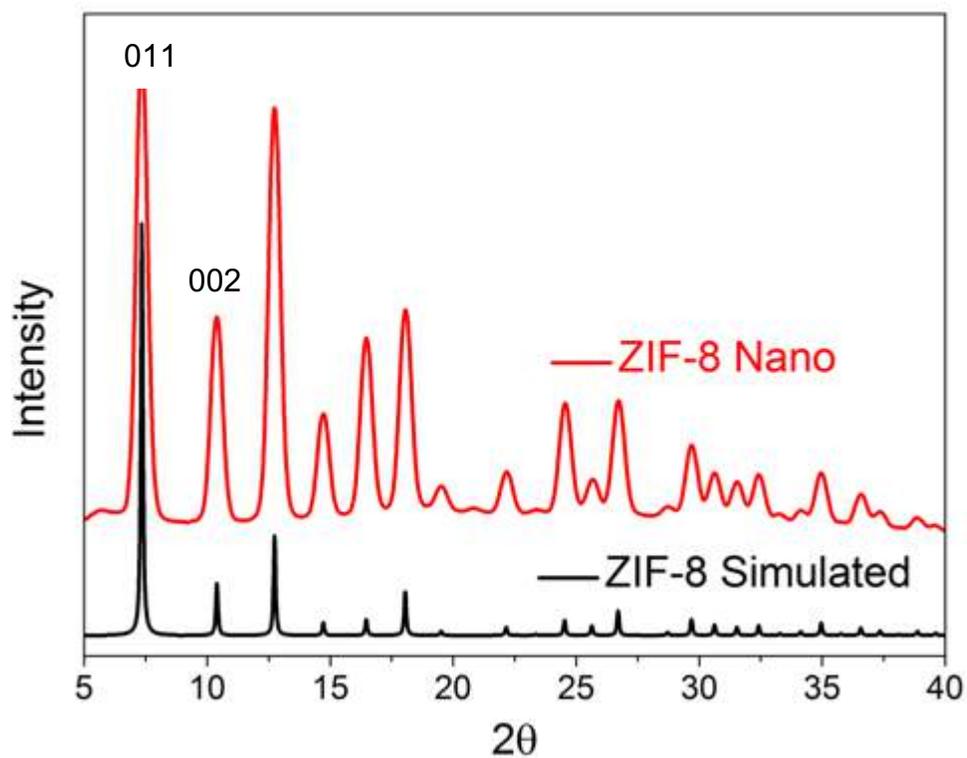

**Fig. S1. Powder XRD profiles.** Simulated (black) and experimental (red) profiles of the body-centered cubic ZIF-8 nanocrystals are plotted, demonstrating that the sample is crystalline and matches the simulated structure.

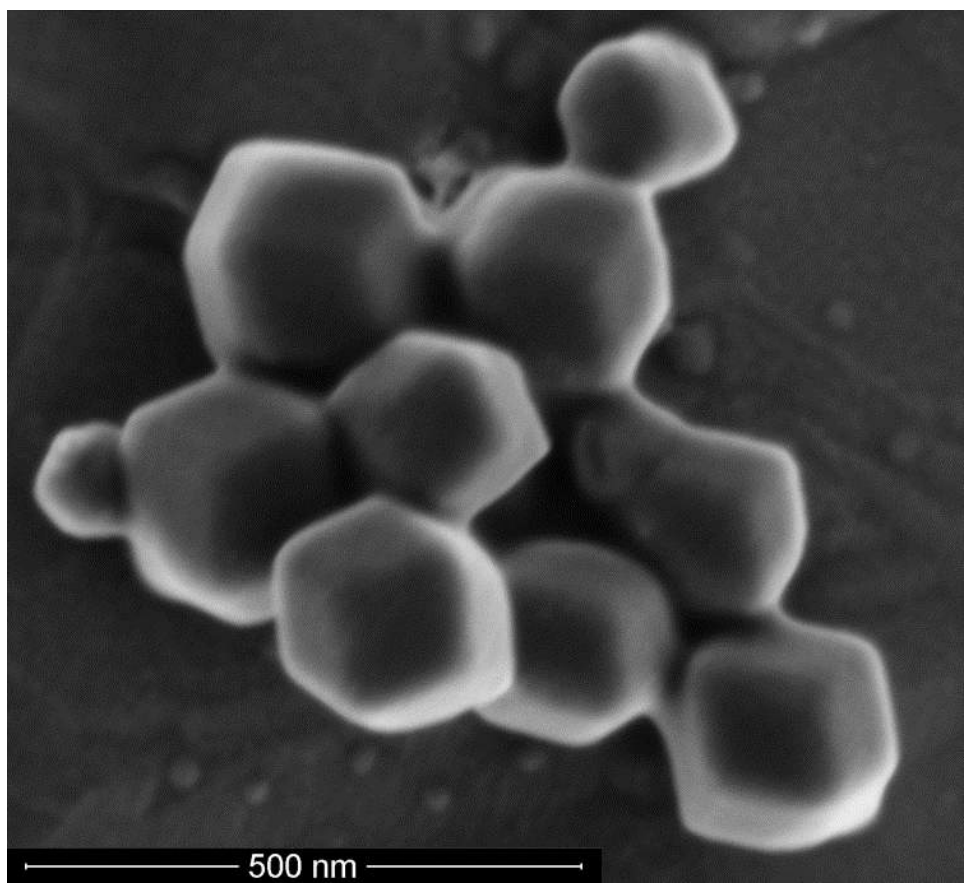

**Fig. S2. SEM of ZIF-8 particles.** The rhombic dodecahedral shape of the ZIF-8 nanocrystals can be observed, with clear faceting at the particle faces. Particle sizes are on the order of 100 nm in thickness. Secondary electron SEM images taken at 1 kV.

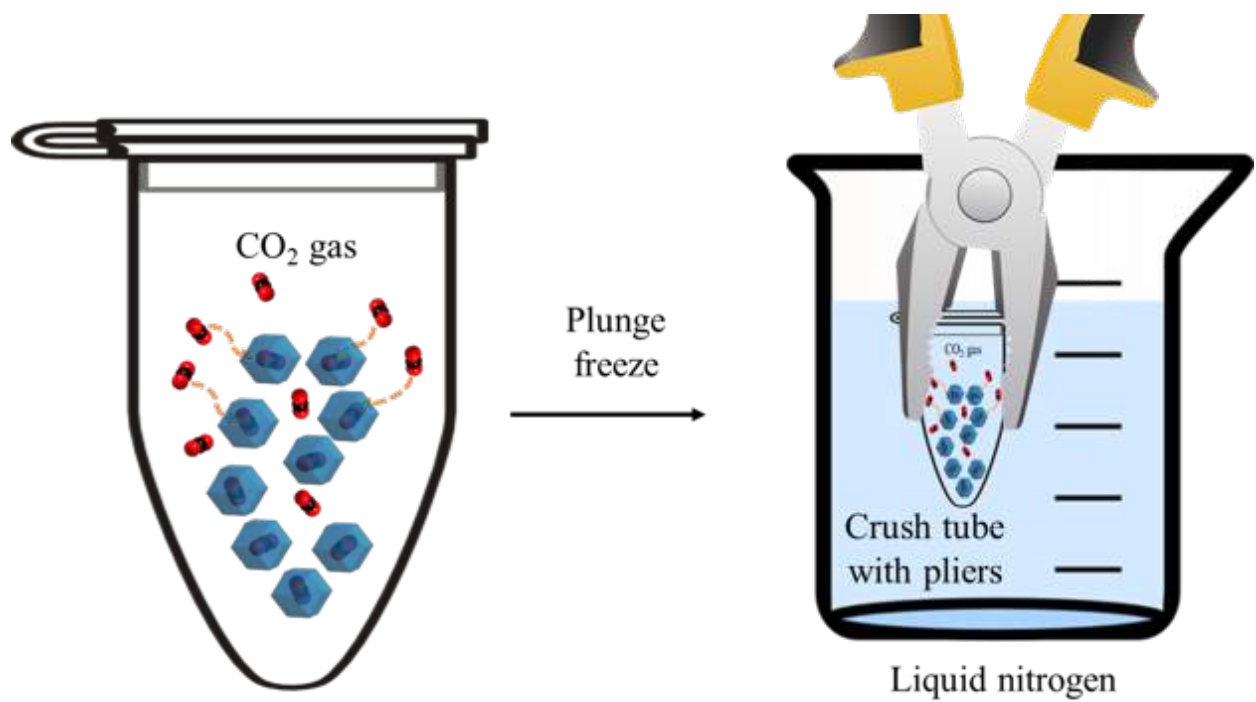

**Fig. S3. Schematic of freezing protocol during CO$_2$ loading.** MOF particles on a TEM grid are placed in an Eppendorf tube filled with CO$_2$ gas. The tube is then plunged into liquid nitrogen and crushed with pliers.

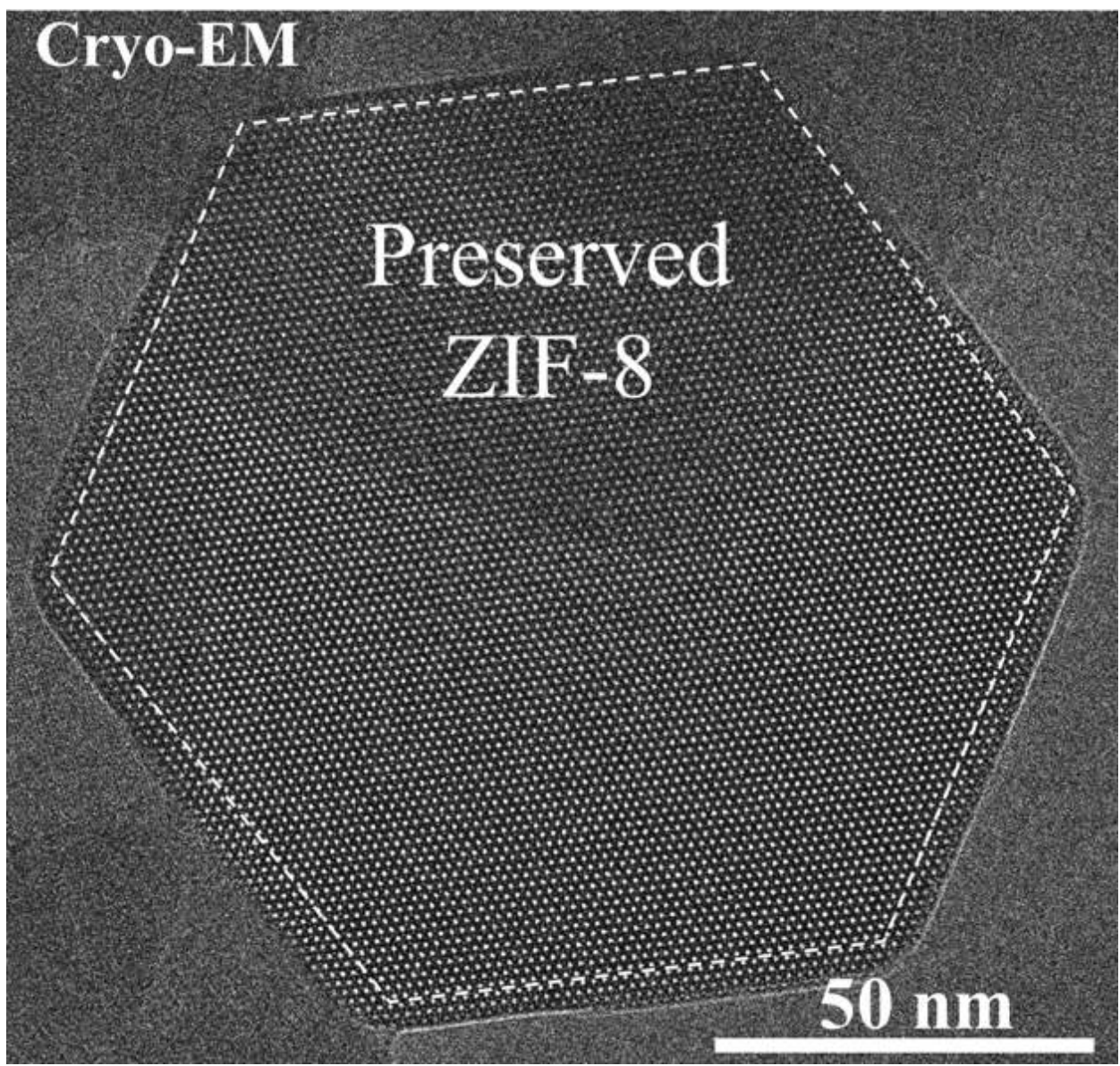

**Fig. S4. Magnified image of Fig. 2c.** Drift-corrected cryo-EM image of ZIF-8 taken at –225 nm defocus along the <111> zone axis. Bright spots indicate pore space within the 6-ring windows. Facets exposed are the {011} family of planes.

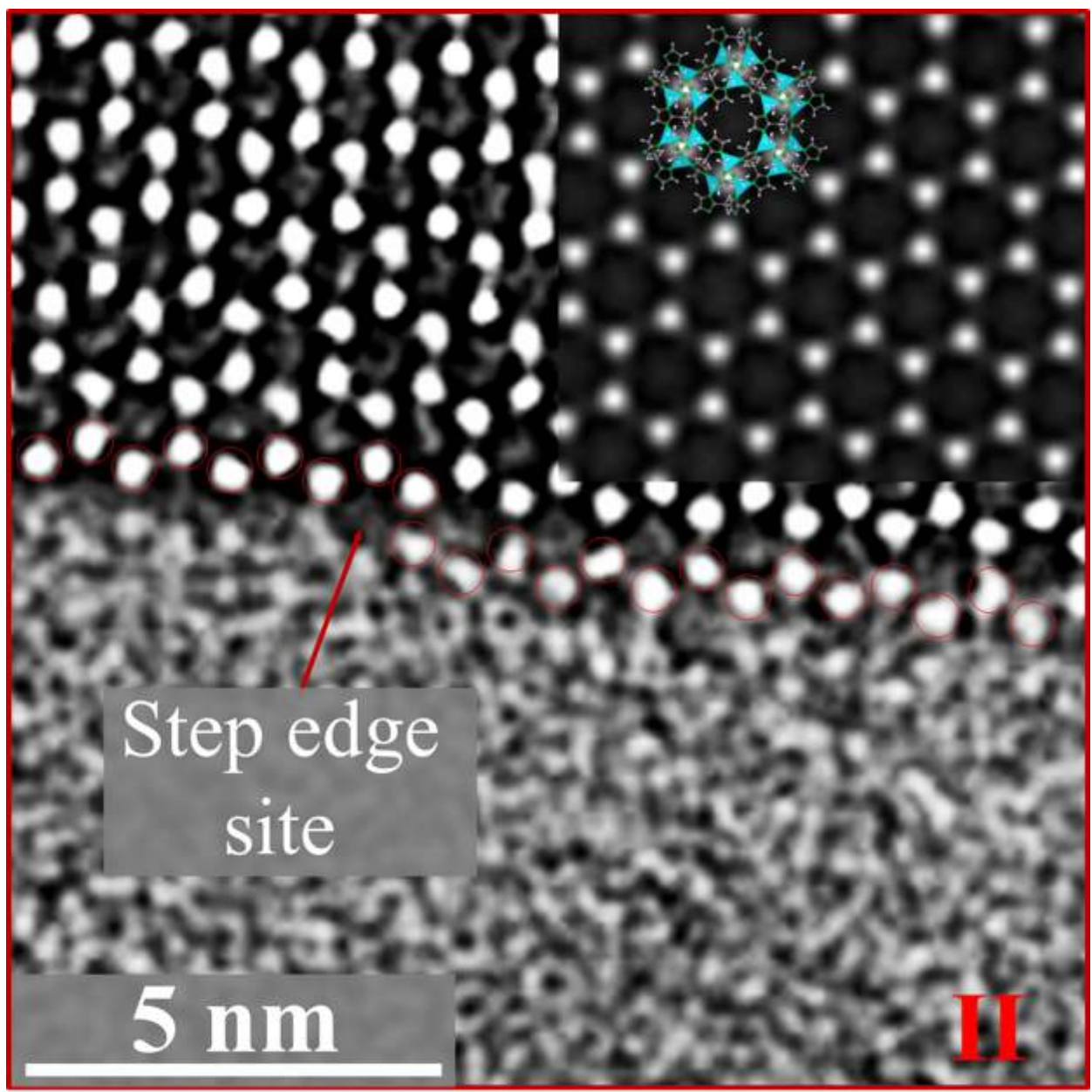

**Fig. S5. Magnified image of Fig. 3c.** Bright spots correspond to Zn metal clusters at positive defocus of 250 nm. The simulated ZIF-8 structure (inset) matches reasonably well with the drift-corrected cryo-EM image. Step edge site is clearly visible on the surface.

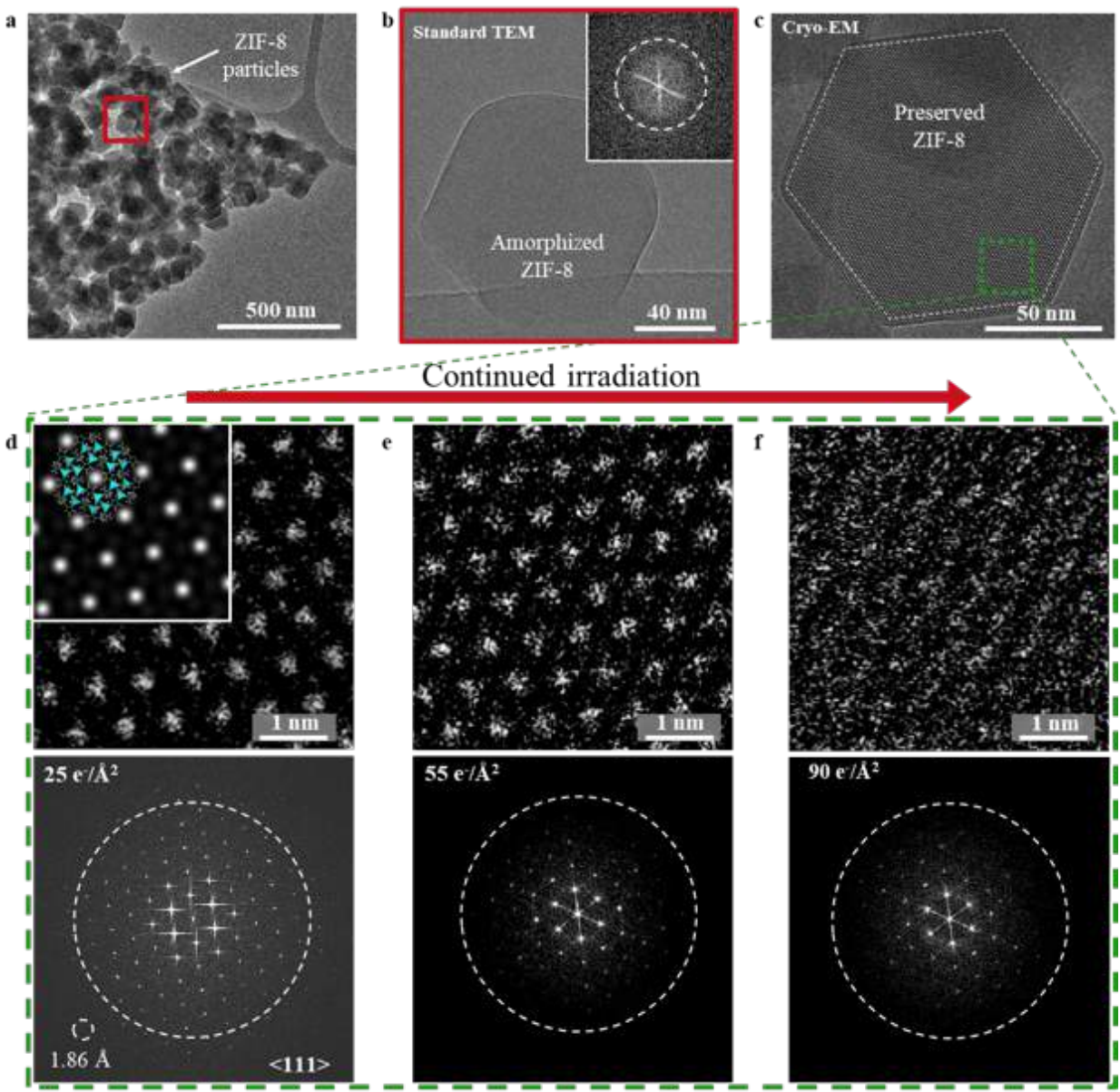

**Fig. S6. Raw, unfiltered, uncorrected images of Fig 2.** Fig. S6c-S6f here are the raw images without any post-processing (i.e. denoising or CTF-correction) corresponding to Fig. 2c-2f in the main text. The inner ring of FFT spots corresponds to the {011} reflections.

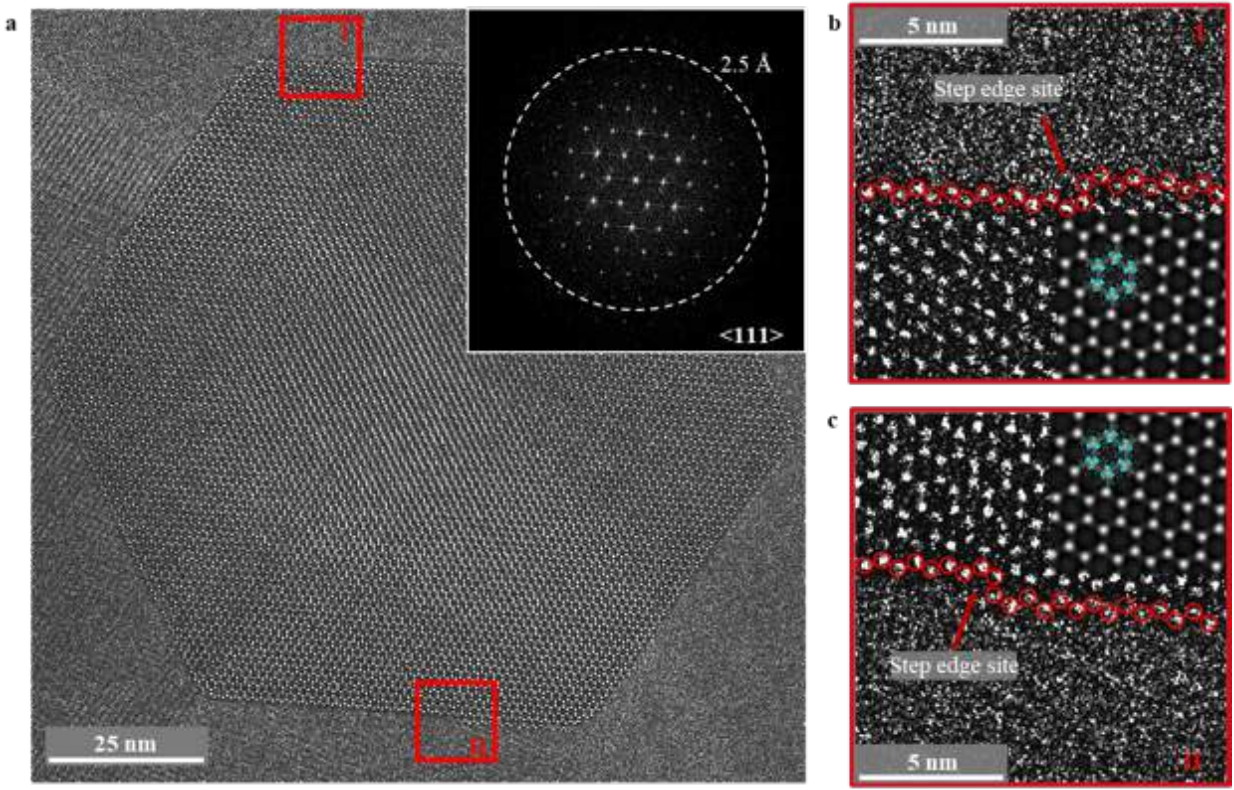

**Fig. S7. Raw, unfiltered, uncorrected images of Fig 3.** Fig. S7a-S7c here are the raw images without any post-processing (i.e. denoising or CTF-correction) corresponding to Fig. 3a-3c in the main text.

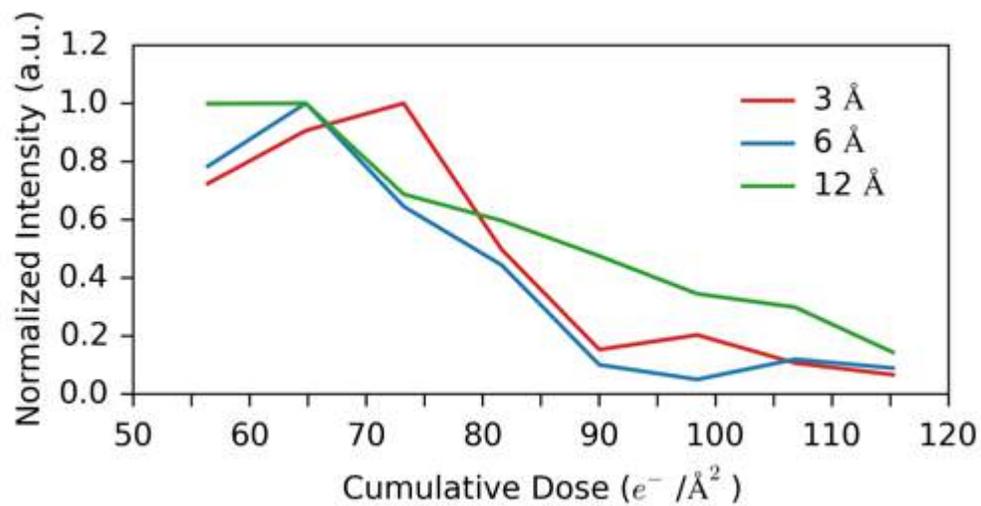

**Fig. S8. Quantitative characterization of irradiation tests.** Plot of the normalized intensity of 3 selected regions from the FFT's in Fig. 2d-2f in the main text after exposure to various cumulative exposures at liquid nitrogen temperature. For example, the red curve (3 Å) represents the region in the FFT that corresponds to 3 Å resolution.

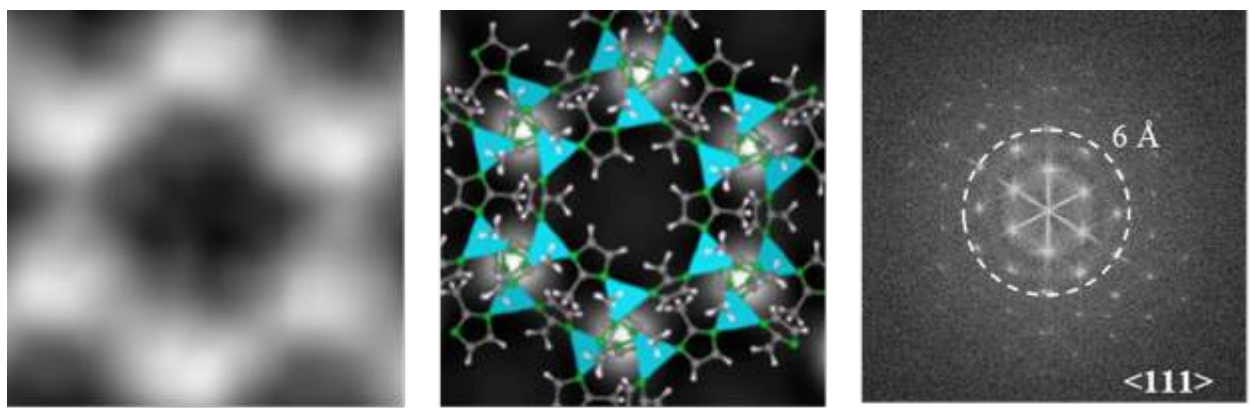

**Fig. S9. HRTEM simulation of empty ZIF-8.** (Left) Experimental filtered image (obtained by masking out all spots in the FFT using the array-mask function from DigitalMicrograph and then using those spots to generate an inverse FFT) of empty ZIF-8 unit cell. (Middle) Simulated image of an empty ZIF-8 unit cell superimposed on the projected structure that matches reasonably well with the experimental image. (Right) Corresponding FFT of the ZIF-8 particle from Fig. 4a after exposure to ~100 $e^-/Å^2$.

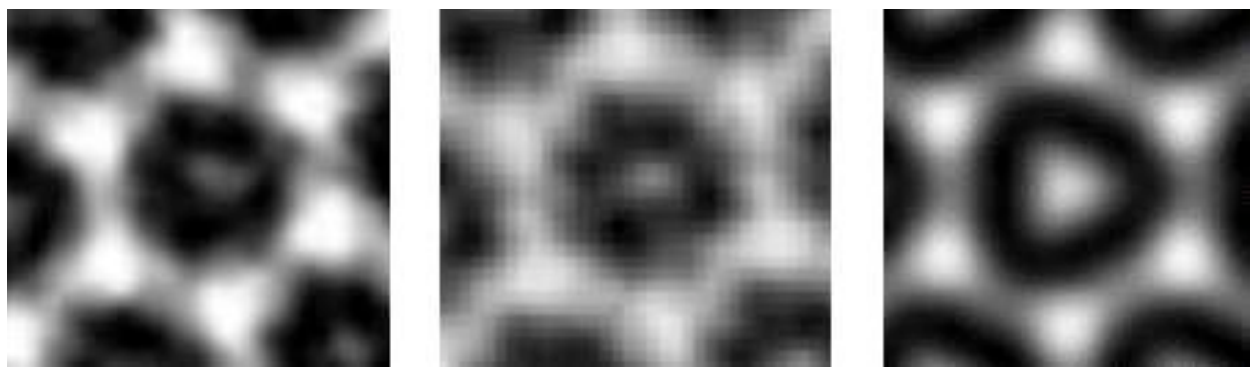

**Fig. S10. HRTEM simulation of $CO_2$-filled ZIF-8.** (Left) Denoised CTF-corrected cryo-EM image of $CO_2$-filled ZIF-8 unit cell. (Middle) Experimental filtered image (obtained by masking out all spots in the FFT using the array-mask function from DigitalMicrograph and then using those spots to generate an inverse FFT) of $CO_2$-filled ZIF-8 unit cell. (Right) Simulated image of $CO_2$-filled ZIF-8 unit cell that matches reasonably well with the experimental image.

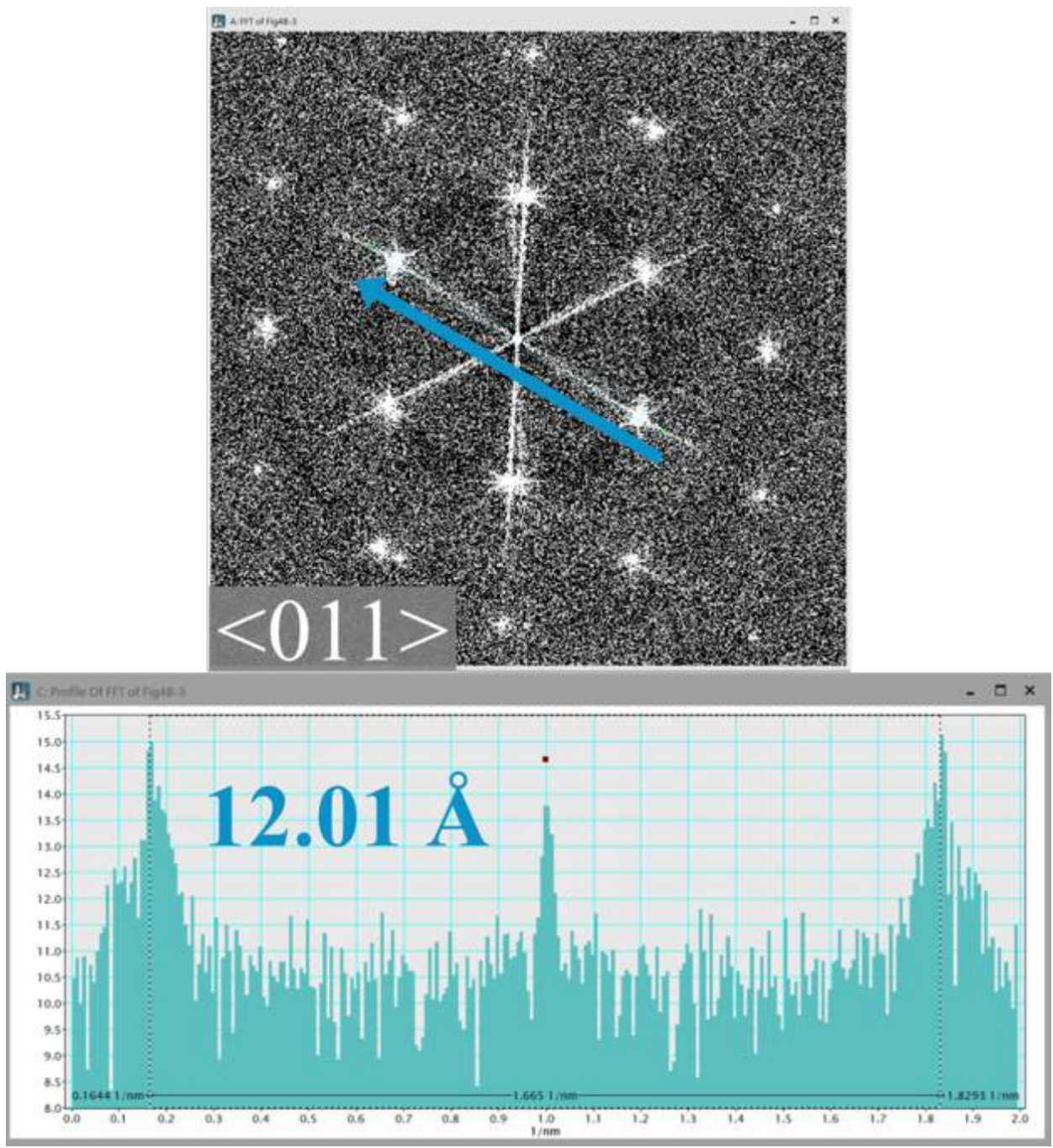

**Fig. S11. Empty ZIF-8 <011> lattice measurement using FFT.** Screenshots of FFT (top) and corresponding intensity plot (bottom) of the empty ZIF-8 particle. The <011> lattice spacing is measured to be 12.01 Å, closely matching the XRD measured value of 12.03 Å. The camera constant was established using a standard Au cross grating TEM grid.

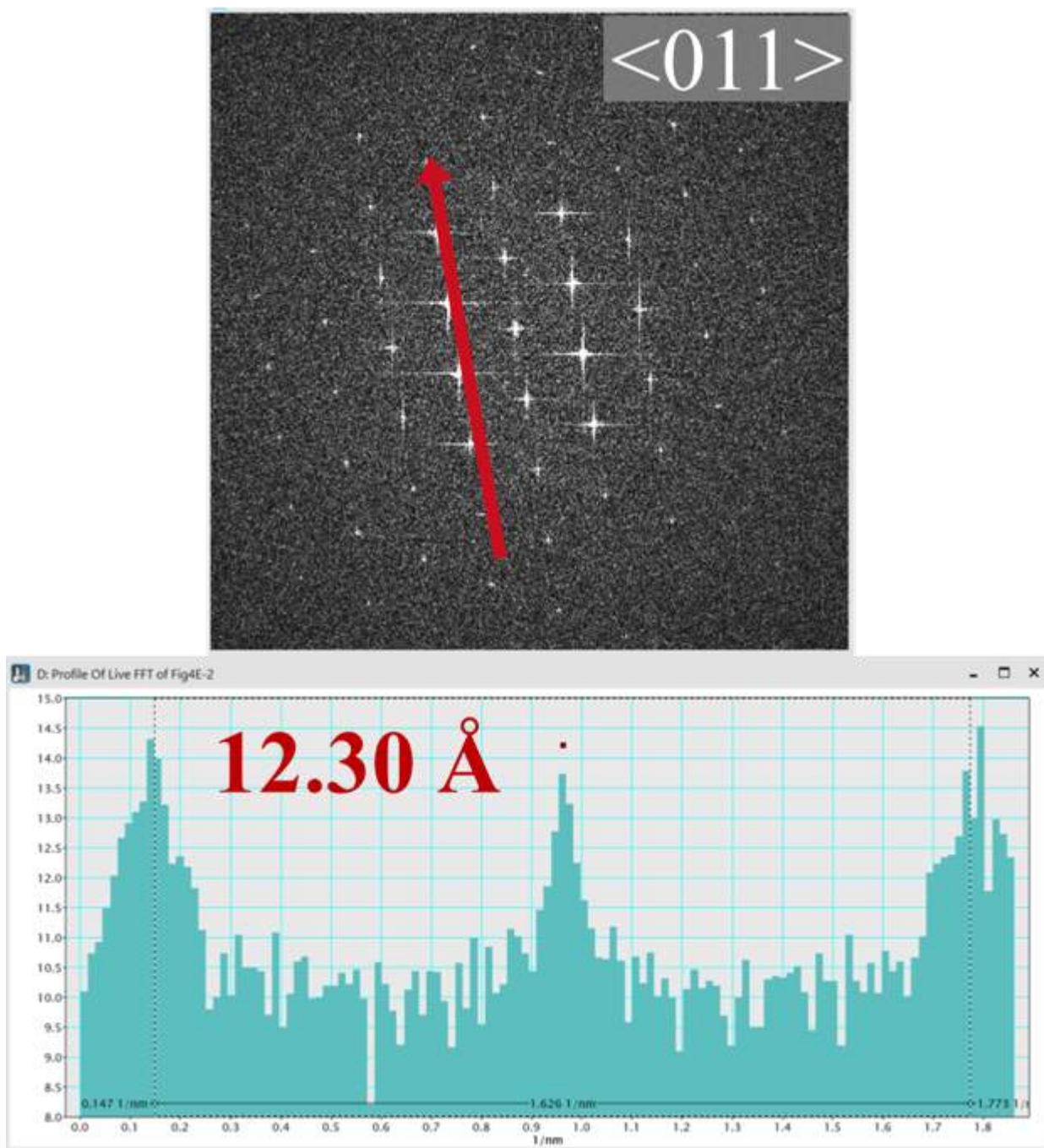

**Fig. S12. CO$_2$-filled ZIF-8 <011> lattice measurement using FFT.** Screenshots of FFT (top) and corresponding intensity plot (bottom) of the CO$_2$-filled ZIF-8 particle. The <011> lattice spacing is measured to be 12.30 Å, an expansion of ~3%.

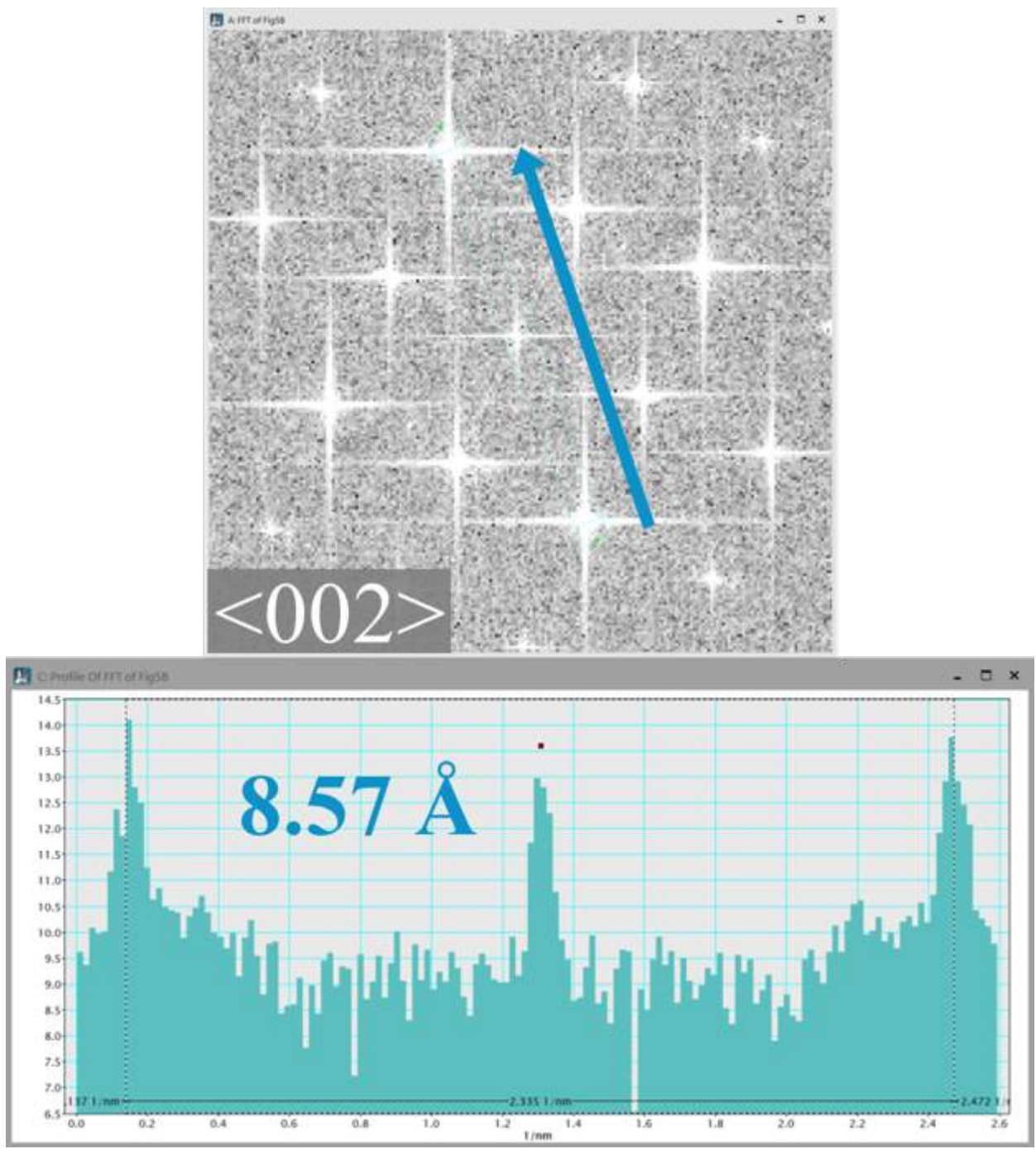

**Fig. S13. Empty ZIF-8 <002> lattice measurement using FFT.** Screenshots of FFT (top) and corresponding intensity plot (bottom) of the empty ZIF-8 particle. The <002> lattice spacing is measured to be 8.57 Å, close to the XRD measured value of 8.50 Å.

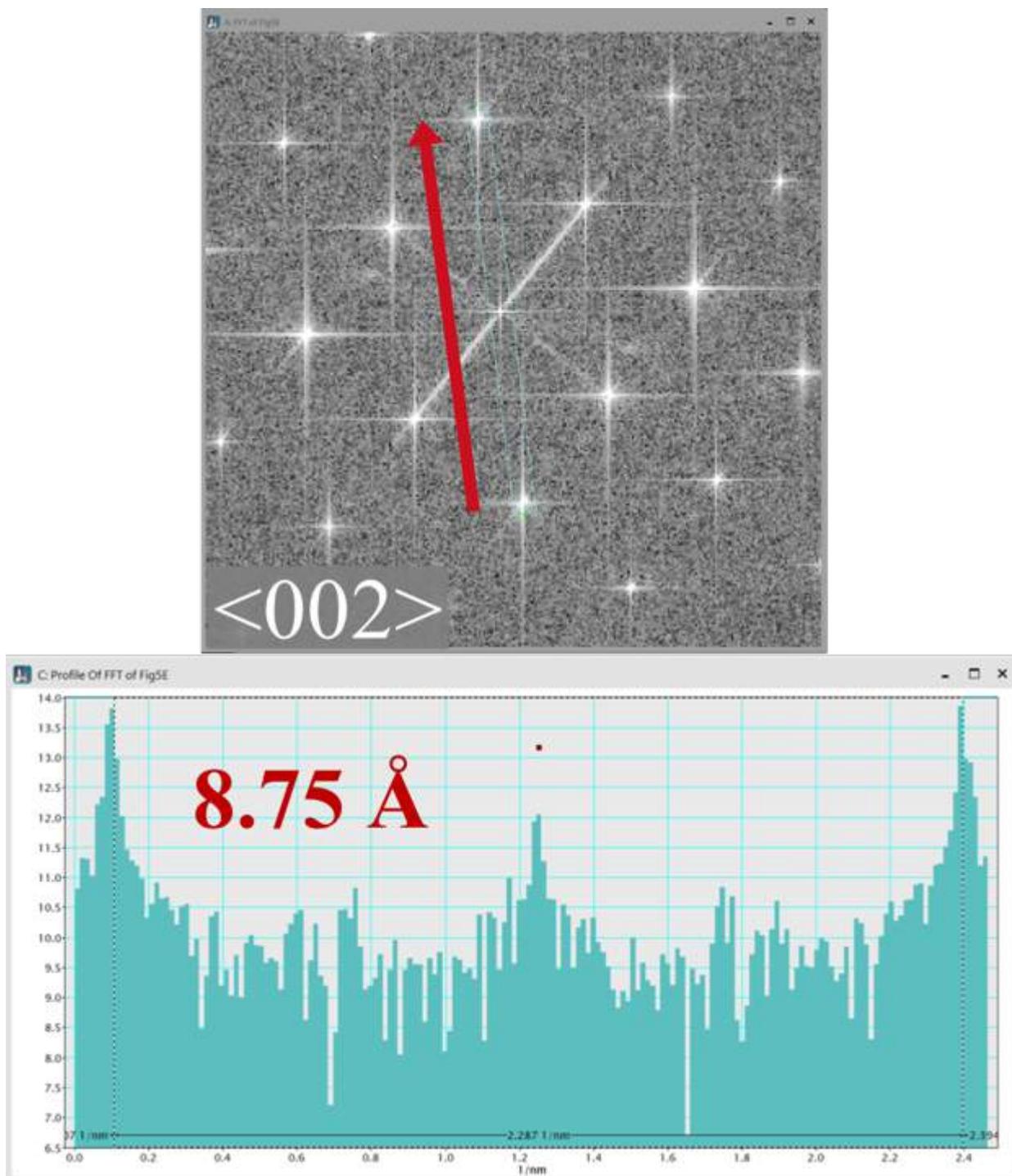

**Fig. S14. CO$_2$-filled ZIF-8 <002> lattice measurement using FFT.** Screenshots of FFT (top) and corresponding intensity plot (bottom) of the CO$_2$-filled ZIF-8 particle. The <002> lattice spacing is measured to be 8.75 Å, an expansion of ~3%.